\documentclass[aps,prb,reprint,superscriptaddress]{revtex4-1}
\usepackage{graphicx}
\usepackage{amsmath}
\usepackage{color}
\usepackage{rotating}
\definecolor{purple}{rgb}{0.7,0.0,0.7}
\definecolor{orange}{rgb}{1,0.65,0.0}

\def\bra#1{\mathinner{\langle{#1}|}}
\def\ket#1{\mathinner{|{#1}\rangle}}

\begin{document}


\title{Electronic states in finite graphene nanoribbons: Effect of charging and defects}


\author{M. Ij\"{a}s}
\email[]{mari.ijas@aalto.fi}
\author{M. Ervasti}
\author{A. Uppstu}
\affiliation{ COMP Centre of Excellence and Helsinki Institute of Physics, Department of Applied Physics, Aalto University School of Science, FI-00076 Espoo, Finland}
\author{P. Liljeroth}
\affiliation{Department of Applied Physics, Aalto University School of Science, PO Box 15100, 00076 Aalto, Finland}
\author{J. van der Lit}
\author{I. Swart}
\affiliation{Condensed Matter and Interfaces, Debye Institute for Nanomaterials Science, Utrecht University, PO Box 80000, 3508 TA Utrecht, the Netherlands}
\author{A. Harju}
\affiliation{ COMP Centre of Excellence and Helsinki Institute of Physics, Department of Applied Physics, Aalto University School of Science, FI-00076 Espoo, Finland}


\date{\today}

\begin{abstract}

We study the electronic {structure} of finite armchair graphene nanoribbons using density-functional theory and the Hubbard model{, concentrating on the states localized at the zigzag termini}. We show that the {energy} gaps between end-localized states are sensitive to doping, and that in doped systems, the {gap between the end-localized states} decreases exponentially as a function of the ribbon length. Doping also quenches the antiferromagnetic coupling between the end{-localized} states leading to a spin-split gap in neutral ribbons. By comparing $dI/dV$ maps calculated using the many-body Hubbard model, its mean-field approximation and density-functional theory, we show that the use of a single-particle description is justified for graphene $\pi$ states. Furthermore, we study the effect of structural defects in the ribbons {on their electronic structure}. Defects at one ribbon termini do not significantly modify the electronic states {localized at} the intact end. This provides further evidence for the interpretation of a multi-peaked structure {in a recent scanning tunneling spectroscopy (STS) experiment} resulting from inelastic tunneling processes [J. van der Lit \emph{et al.}, Nature Commun., in press (2013)]. Finally, we show that the hydrogen termination at the flake edges leaves identifiable fingerprints on the positive bias side of STS measurements, thus possibly aiding the experimental identification of graphene structures.  
\end{abstract}


\maketitle


\section{Introduction}

At zigzag-terminated graphene edges, the presence of non-dispersive, edge-localized states was predicted {nearly 20 years ago} based on the tight-binding model.\cite{Nakada} Taking electron-electron interaction into account by using density-functional theory (DFT) \cite{Son} or the Hubbard model in the mean-field level \cite{Jung} opens a gap between the doubly degenerate edge-localized states, {with a} magnitude {that} is inversely proportional to the ribbon width. Moreover, the edges show antiferromagnetic order {with respect to each other}, with the electrons belonging to the two spin channels predominantly occupying opposite edges, {and ferromagnetic order within a given edge}.\cite{Son} Recently, the magnetic correlations in finite armchair nanoribbons were studied in detail using an effective low-energy model.\cite{Golor} Thus far, only few observations of this spin-split {gap} have been reported in large-scale graphene nanoribbons, for which the structure of the edge is not known with atomic detail.\cite{Joly, Tao, Pan} 

Using top-down methods, such as etching or unzipping of carbon nanotubes, the preparation of graphene nanoribbons with atomically well-defined edges is challenging. Recently, a bottom-up approach based on the on-surface polymerization of 10,10'-dibromo-9,9'-bianthryl precursors was introduced.\cite{Cai} The advantage of this approach is that the structure of the precursor molecule determines also the edge termination of the resulting ribbon. Thus, {the synthesis of} seven carbon rows wide armchair nanoribbons (7-AGNRs) with zigzag-terminated ends is well established. Also double- and triple-width ribbons have been reported.\cite{Huang} The electronic structure of these ribbons has been widely studied,\cite{Ruffieux, Talirz, Koch, Linden, vanderLit, Bronner} finding a bulk band gap of 2.3-5.1~eV using scanning tunneling spectroscopy (STS) \cite{Ruffieux,Talirz,Koch,Linden,vanderLit}, angle-resolved photoelectron spectroscopy (ARPES)\cite{Ruffieux}, and optical methods.\cite{Linden,Bronner} Within the bulk gap, states localized at the ribbon zigzag ends have been observed.\cite{Bronner,Talirz,vanderLit} A double-peak structure resembling the spin-split zigzag end states was indeed recently observed.\cite{vanderLit} {It was, however, found to arise from phonon-assisted tunneling, since the side peak energies agreed well with ribbon phonon frequencies, and the peak shapes agreed with theoretical predictions for phonon-assisted tunneling.

Computational studies have accompanied the experiments on the surface-deposited ribbons but the {end-localized electronic states} in the finite ribbons has not been thoroughly addressed. {The effect of ribbon length} as well as that of doping caused by the substrate on the {spin-split gap and ribbon electronic structure, in general,} remains unknown. In the scanning tunneling microscopy experiments, states with different spatial $dI/dV$ maps have been observed at the ribbon ends, and these states have been attributed to different hydrogen terminations at the ends by comparing experiments to computational simulations.\cite{Ruffieux, Koch, Talirz} The effect of structural imperfections at the armchair edges, experimentally introduced by applying voltage pulses,\cite{Talirz,vanderLit} on the low-bias electronic structure and $dI/dV$ measurements is yet to be addressed. 

In this paper, we study the electronic states {in finite 7-AGNRs, concentrating on states} localized at the zigzag termini. Using the Hubbard model and density-functional theory, we address the role of correlation phenomena in neutral and charged flakes, as well as justify the use of Kohn-Sham energy levels and wavefunctions to model STS. Furthermore, in relation to experiments presented in Ref.~\onlinecite{vanderLit}, we show that modifying one ribbon end does not alter {the electronic states} {localized at} the intact end. {This supports the conclusion of Ref.~\onlinecite{vanderLit} that the peak structure arises from phonons}. Finally, we simulate $dI/dV$ maps for ribbons with different edge hydrogenation patterns, and identify their fingerprints in the $dI/dV$ maps. 

\section{Computational methods}

The modeled {7-AGNR} structures consist of three to nine monomer units, their length {thus} ranging from approximately 2.5 nm to 7.5 nm. Fig.~\ref{fig:TB_edge_orbitals} shows the 28-carbon atom monomer unit, marked by a gray background {in the three-monomer ribbon}. As the interaction between the Au(111) surface and the ribbons is weak,\cite{vanderLit} we model freestanding ribbon fragments and take the presence of the substrate into account only through doping.\cite{Archambault} In addition to doping, the substrate is also expected to screen electron-electron interactions.\cite{Hwang} {Including the substrate in the calculation would, however, be computationally extremely demanding, especially for longer ribbons.}

{We use both the Hubbard model\cite{Hubbard} and density-functional theory (DFT) to study freestanding finite 7-AGNRs.} The Hubbard Hamiltonian {for the $\pi$-electrons}  is defined on the lattice {formed by the carbon atoms} as
\begin{equation} \label{eq:Hubbard} H = \sum_{\sigma; i,j} t_{ij} c_{i\sigma}^{\dagger}c_{j\sigma} + U \sum_i n_{i\uparrow}n_{i\downarrow},
\end{equation}
where $i$ and $j$ denote the lattice sites, $t_{ij}$ are the tight-binding hopping elements between sites $i$ and $j$, $U$ is the on-site repulsion, {$c_{i \sigma}$ annihilates a spin-$\sigma$ electron from site $i$,} and $n_{i\sigma} = c_{i\sigma}^{\dagger}c_{i\sigma}$ is the {site} occupation operator. The tight-binding model corresponds to Eq.~(\ref{eq:Hubbard}) with $U=0$. We use the parameters $t_1 = -2.7$~eV, $t_2 = -0.2$~eV, and $t_3 = -0.18$~eV, {between sites that are first, second, and third-nearest neighbors}, and for the Hubbard on-site interaction parameter the value $U = 2$~eV.\cite{Uppstu} 

The many-body calculation utilizes the Lanczos algorithm to solve $\sim100$ lowest {many-body} eigenstates accurately. The state space is formed from a given symmetry sector by constructing many-body configurations of tight-binding orbitals, {and by ordering them according to the} energy of the tight-binding part of the Hamiltonian. Of these configurations, roughly 10$^6$ lowest in energy are included to the many-body basis used in the calculation. This constitutes an orbital-dependent approximation with fast convergence. Due to the associated computational effort, {we focused on a three-monomer ribbon using this approach.}

In the mean-field {approximation}, the two-body interaction is reduced to a one-body potential {that is} determined by the  electron density of the opposite-spin electrons on each site,
\begin{equation} n_{i\uparrow}n_{i\downarrow} \mapsto \langle n_{i\uparrow} \rangle n_{i\downarrow} + n_{i\uparrow} \langle n_{i\downarrow}\rangle - \langle n_{i\uparrow} \rangle \langle n_{i\downarrow} \rangle,
\end{equation}
where $\langle n_{i \sigma} \rangle$ denotes the site occupation. The resulting mean-field Hamiltonian is solved self-consistently {for both spin components,} until the electron densities and ground-state energy have converged.

The DFT calculations were performed using the all-electron code "FHI-aims".\cite{AIMS} The "tight" basis defaults for numeric atom-centered orbitals, as defined in the FHI-aims distribution, were used{. In} calculations using a hybrid functional,  the "light" defaults were chosen to reduce computational cost. The structures were relaxed using the Perdew-Burke-Ernzerhof (PBE)\cite{PBE} exchange-correlation functional until the total forces acting on atoms were less than 10$^{-3}$~eV/\AA{}, and the total energy was converged to 10$^{-6}$~eV. The spin initialization used was that of a neutral, fully hydrogenated ribbon, with antiferromagnetic coupling between the ribbon ends. The PBE exchange-correlation functional was used unless otherwise specified. Additionally, some structures were calculated with the B3LYP hybrid functional to find out the effect of the inclusion of explicit Hartree-Fock exchange, improving the description of electron-electron interactions. In calculations with a non-zero $z$-component of the {total} spin, $S_z=(N_{\mathrm{el},\uparrow}-N_{\mathrm{el},\downarrow})/2$, the number of spin up and spin down electrons was fixed. 

Differential conductance maps were simulated based on the Tersoff-Hamann model,\cite{Tersoff-Hamann} according to which the $dI/dV$ signal of a $s$-wave tip is proportional to the local density of states $\rho(\vec{r}, E)$ (LDOS) in the sample, 
\begin{align} \label{eq:TH} 
\frac{dI(\vec{r}, V)}{dV} &\propto \rho(\vec{r}, E_F+eV) \nonumber \\
&=\sum_n \left|\Psi_n(\vec{r})\right|^2 \delta(E_F+eV-E_n).
\end{align}
Here, $V$ is the bias voltage, and ($\Psi_n$, $E_n$)  are the molecular orbitals  and {the} corresponding eigenenergies. {Experimental $dI/dV$ peaks} are broadened in energy due to the weak interaction with the substrate, temperature, and instrumental precision. Thus, the energy delta function is broadened into a Lorentzian, 
\begin{equation}
\label{eq:broaden}
\delta(\epsilon) =\frac{1}{\pi} \frac{\eta}{\epsilon^2+\eta^2},
\end{equation}
where $\eta$ is the broadening parameter, chosen to be 50~meV in the present study. The exact value mainly affects the width of the peaks in the density of states (DOS) figures, and has only little effect on the simulated $dI/dV$ maps.  

{{In the STS experiments, the current flowing between the substrate and the tip is measured. Depending on the applied bias voltage, the tunneling occurs either from the tip to the substrate or vice versa. In general, STM and STS probe the hybrid substrate-molecule system, and the signal is also dependent on the tip characteristics, as well as the strength of the coupling to the tip that is assumed to be weak. For a {weak enough coupling between the substrate and the molecule}, the lifetime of the charge carrier on the molecule is long, and the molecular orbitals are probed. At negative and positive bias, the molecule is temporarily hole-doped and electron-doped, respectively.} }

{In theoretical modeling based on DFT or other effective single-electron models,  the molecular orbitals of the $N$-electron system are used to model the measurements both on the negative and positive bias sides (occupied and unoccupied orbitals, respectively). {The change of electron occupation is not taken into account.}
Moreover, these are single-particle {orbitals}, whereas the experiment actually probes transitions between many-body states. In order to evaluate the validity of this approach, it is interesting to compare the simulated DOS and $dI/dV$ maps {given by} the effective one-body descriptions, DFT and the mean-field Hubbard model, to the spectral function {given by} a full many-body calculation that includes the effects due to the changing electron occupation. }

The local density of states is generalized to the many-body picture by the spectral function, whose diagonal elements are defined for $N=(N_\uparrow, N_\downarrow)$ particles at zero temperature as
\begin{align} \label{eq:specfunc}
& A(\nu \sigma; \omega) = -2 \: \text{Im} [ \: G^R(\nu\sigma; \omega)]  \notag\\
= &\sum_{k} \left\| \bra{\Psi_k^{(N_\sigma+1)}} c_{\nu\sigma}^\dagger \ket{\Psi_0^{(N)}} \right\|^2 \: 2 \pi \delta \left( \omega - E_k^{(N_\sigma+1)} + E_0^{(N)} \right) \notag\\
+&\sum_{k} \left\| \bra{\Psi_k^{(N_\sigma-1)}} c_{\nu\sigma} \ket{\Psi_0^{(N)}} \right\|^2 \: 2 \pi \delta \left( \omega + E_k^{(N_\sigma-1)} - E_0^{(N)} \right) ,
\end{align}
where $\ket{\Psi_k^{(N)}}$ is the $k$:th {many-body} eigenstate with energy $E_k^{(N)}${, and $\nu$ ($\sigma$) is the single-electron state (spin) index}. In {the} position basis, the spectral function diagonal corresponds to {the LDOS}, and to $dI/dV$ maps in tunneling spectroscopy. In corollary, tracing over the single particle orbitals $\nu$ produces {an} equivalent {of}  the density of states {multiplied by a factor of $2\pi$}. In fact, the non-interacting case reduces to Eq. \eqref{eq:TH}.

The original Tersoff-Hamann model considers only spherically symmetric $s$-wave tips.\cite{Tersoff-Hamann} The model was later extended to other tip symmetries, such as $p$-wave tips, and the extension was formulated in terms of a simple derivative rule.\cite{Chen,Gross} Writing the tunneling matrix element in Eq.~(\ref{eq:TH}) as $M_s = \left| \Psi_n(\vec{r})\right|^2$, the corresponding matrix elements for $p_x$ and $p_y$ type tips are given by $M_{p_x}(\vec{r}) = \left| \partial \Psi(\vec{r}) / \partial x\right|^2$ and $M_{p_y}(\vec{r}) = \left| \partial \Psi(\vec{r}) / \partial y\right|^2$. A cylindrically symmetric CO tip is obtained by combining $M(\vec{r}) = M_{p_x}(\vec{r})+M_{p_y}(\vec{r})$, and tips mixing both $s$ and $p$ character are also possible.\cite{Gross}  

In the experiment, the distance between the sample and the tip can be rather large, up to 10~\AA{}. In the DFT calculation, the atom-centered numeric basis set has to be cut off at some distance, and due to excessive computational effort, equally large distances {cannot} be achieved. {In} the simulated $dI/dV$ maps, a height of 3.5~\AA{} was chosen. {The results {are} qualitatively insensitive to the exact height, as long as it {is} large enough {{to exclude most of} the $\sigma$-type orbital contribution}  forming the in-plane carbon-carbon bonds. A minimal height of 2~\AA{} was found to be sufficient in the current DFT calculation.}  As the Hubbard model is defined on a set of lattice sites, {the lattice basis has to be transformed into {a basis of} real space {orbitals} in order to evaluate {the} LDOS at position $\vec{r}$ {for} the $dI/dV$ maps. An analytical carbon $p_z$ orbital\cite{Radzig} {is placed} onto each lattice site, and the LDOS {is evaluated} at {a} tip height of 4.0~\AA{} in the Hubbard calculations.} The gray scale of the $dI/dV$ maps {in the figures} has been normalized separately in each map, ranging from white (highest intensity) to black (no intensity). {T}he associated DOS plots {can be used to compare} the magnitudes of the simulated maps.

Unless otherwise specified, {in the DFT calculations} the position of zero energy has been fixed to the middle of the gap between the highest occupied molecular orbital (HOMO) and lowest unoccupied molecular orbital (LUMO). This energy is referred to as $E_{\mathrm{ref}}$.

\section{Results}

\subsection{Neutral hydrogenated ribbon}

\begin{figure}
\includegraphics[scale=0.5]{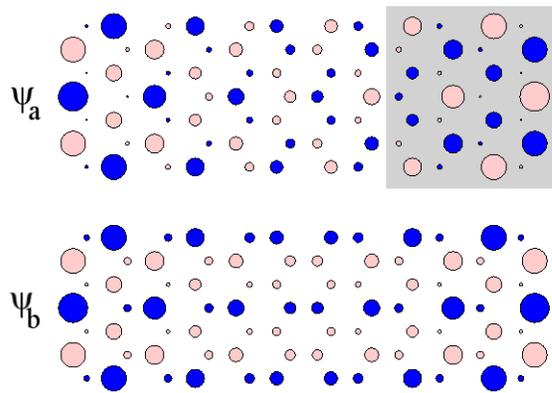}
\caption{\label{fig:TB_edge_orbitals} (Color online) {The nearly degenerate end-localized orbitals, $\psi_a$ and $\psi_b$, from a tight-binding calculation. Their energy difference is $\approx$0.05~eV and it decreases with an increasing ribbon length. Both figures are on the same scale, and the area of the lattice sites correspond to the amplitude of the wavefunction (the maximum value 0.323), and the {light red/blue (gray/dark)} color {indicates} the sign of the wavefunction. The gray background marks the lattice sites belonging to one monomer unit. }}
\end{figure}

\begin{figure*}
\includegraphics[width = 1.9\columnwidth]{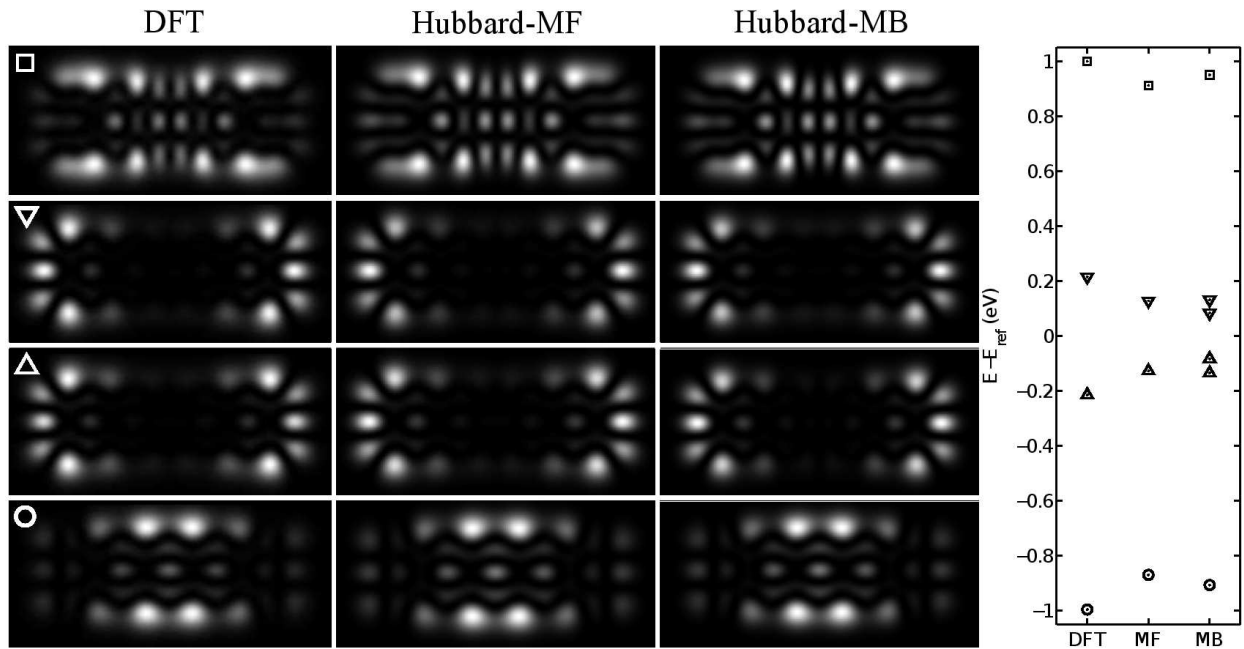}
\caption{\label{fig:methodcomp} A comparison between DFT, mean-field (MF) Hubbard model and the many-body  (MB) Hubbard model used for simulating the $dI/dV$ maps {of a three-monomer ribbon} with a $s$-wave tip. The maps have been calculated at the molecular orbital energies (DFT, mean-field Hubbard) or at the peak positions of the many-body spectral function. On the right, peak energies are indicated. In the many-body spectra, the energy reference has been set to the middle of the first peaks obtained by creation and annihilation. {The different markers link the $dI/dV$ maps and corresponding energies.}}
\end{figure*}

As the tight-binding and Hubbard models assume passive, hydrogen-terminated edges, we first consider a neutral fully hydrogenated flake {and compare the simulated $dI/dV$ maps to results obtained using DFT}. {At half-filling, the tight-binding model  (corresponding to Eq.~(\ref{eq:Hubbard}) with $U=0$) predicts two eigenstates localized at the zigzag-terminated ribbon ends that are nearly degenerate, and occur on both sides of the Fermi energy $E_F$. } In fact, in the limit of long ribbons, the {energy} gap between the {end} states vanishes, whereas the gaps to states below and above them stay finite ($0.580$~eV and $0.688$~eV{, respectively}). {As the end-localized states lie close to each other in energy, even a weak interaction {results in} correlations.

The two {end states}, denoted $a$ and $b$ with wave functions $\psi_a$ and $\psi_b$, are illustrated in Fig. \ref{fig:TB_edge_orbitals} for the {three-monomer} 7-AGNR. Regardless of the ribbon length, {$\psi_a$ and $\psi_b$} are localized at the zigzag ends, {having} the same exponential decay towards the ribbon center. {In state $\psi_a$, the ends are of opposite sign, whereas in $\psi_b$ they are of the same sign.} This motivates to write $\psi_a=\frac{1}{\sqrt{2}} ( \psi_l -\psi_r ) $ and $\psi_b=\frac{1}{\sqrt{2}} ( \psi_l + \psi_r )$, where {$\psi_{l}$ and $\psi_r$} are {states localized at the left and right end of the ribbon, {respectively}, and each one {resides} mainly on one of the two sublattices}.

The lowest eigenstates of the many-body Hubbard model at half filling{, spanned by the {end} orbital occupations,} can be approximately written as
\begin{align}
\ket{0_{S=0}} &= g^\dagger \left[ \alpha c^\dagger_{a\uparrow} c^\dagger_{a \downarrow} - \beta c^\dagger_{b \uparrow} c^\dagger_{b \downarrow} \right] \ket{0} , \label{eq:MB_GS_S0}\\
\ket{0_{S=1}} &= g^\dagger c^\dagger_{a\uparrow} c^\dagger_{b \uparrow}  \ket{0}, \label{eq:MB_GS_S1}\\
\ket{1_-} &= g^\dagger \frac{1}{\sqrt{2}} \left[ c^\dagger_{a \uparrow} c^\dagger_{b \downarrow} + c^\dagger_{b \uparrow} c^\dagger_{a \downarrow} \right] \ket{0} , \\
\ket{1_+} &= g^\dagger \left[ \beta c^\dagger_{a\uparrow} c^\dagger_{a \downarrow} + \alpha c^\dagger_{b \uparrow} c^\dagger_{b \downarrow} \right] \ket{0} ,
\end{align}
where $c_{a\sigma}^\dagger$ {and} $c_{b\sigma}^\dagger$ create the {end} orbitals $a$ and $b$, {and} $g^\dagger$ creates the frozen core of the lower bulk orbitals, and {the coefficients} $\alpha, \beta \geq 0$. These states approximate the exact eigenstates in the sense that, in the numerical calculations, the other configurations have only minimal weights. For the {three-monomer} 7-AGNR, for instance,  the coefficients are $\alpha^2 \approx 0.69$, $\beta^2 \approx 0.28$, and all the others have a squared weight of $\approx 0.02$ in total. Since the energy gap between the {end} orbitals $\psi_a$ and $\psi_b$ decreases as a function of ribbon length, also $\alpha, \beta$ balance and {seem to} {converge to a common value. {For instance, in a five-monomer ribbon, $\alpha^2 \approx 0.51$ and $\beta^2 \approx 0.48 $.  }

The exact eigenstates corresponding to $\ket{0_{S=0}}$ and $\ket{0_{S=1}}$ are the two lowest in energy, and they are almost degenerate. {They have total spins $S=0$ and $S=1$, as embedded in the notation.} {The other two states $\ket{1_\pm}$ are also almost degenerate, but {their energies are} significantly higher than {that of} the ground state. They have a total spin of $S=0$.} For the three-monomer 7-AGNR, the exact state {corresponding to} $\ket{0_{S=0}}$ is the ground state. However, with increasing ribbon length, both states {($\ket{0_{S=0}}$ and $\ket{0_{S=1}}$)} converge in a similar manner and seem to approach each other in energy. It is also worth noting that without the second-nearest-neighbor couplings {($t_2$)}, Lieb's theorem\cite{Lieb} would imply a ground state {with} $S=0$ due to the sublattice balance of the bipartite lattice.

The magnetic properties {at the ribbon ends} are more easily grasped if the states {$\psi_a$ and $\psi_b$} are written in the basis of the left- and right-localized orbitals {$\psi_l$ and $\psi_r$}. The transformation is given by $c_l^\dagger = \frac{1}{\sqrt{2}} ( c_a^\dagger + c_b^\dagger )$, and $c_r^\dagger = \frac{1}{\sqrt{2}} ( - c_a^\dagger + c_b^\dagger )$. The eigenstates are then given by
\begin{align}
\label{eq:apprgs}\ket{0_{S=0}} &= - g^\dagger \frac{1}{2}  \left[ (\alpha + \beta) ( c^\dagger_{l \uparrow} c^\dagger_{r \downarrow} + c^\dagger_{r \uparrow} c^\dagger_{l \downarrow}) \right.  \notag\\
&+ \left. (\beta-\alpha) (c^\dagger_{l\uparrow} c^\dagger_{l \downarrow} + c^\dagger_{r \uparrow} c^\dagger_{r\downarrow}) \right] \ket{0} , \\
\ket{0_{S=1}} &= g^\dagger c^\dagger_{l\uparrow} c^\dagger_{r\uparrow} \ket{0}\\
\ket{1_-} &= g^\dagger  \frac{1}{\sqrt{2}}[c^\dagger_{l \uparrow}c^\dagger_{l \downarrow}-c^\dagger_{r \uparrow}c^\dagger_{r \downarrow}] \ket{0}\; \mathrm{{and}} \\
\ket{1_+} &= g^\dagger\frac{1}{2}\left[(\alpha-\beta)(c^\dagger_{l \uparrow}c^\dagger_{r \downarrow}+c^\dagger_{r\uparrow}c^\dagger_{l \downarrow})\right. \notag \\
&+\left .(\alpha+\beta)(c^\dagger_{l \uparrow}c^\dagger_{l \downarrow}+c^\dagger_{r \uparrow}c^\dagger_{r \downarrow})
 \right] \ket{0}. 
\end{align}
{The state $\ket{0_{S=0}}$ is antiferromagnetic {across} the two {ribbon} {ends}, as the first term with coefficient $(\alpha+\beta)$ dominates.} Namely, measuring a spin-up particle at the left {end} results in spin-down particles more likely being found at the opposite {end}, and vice versa. Similarly, $\ket{0_{S=1}}$ is ferromagnetic across the two {ends}. The magnetic properties are related to the energetics, {as} the Hubbard interaction energy is larger for states of type $c_{l \uparrow}^\dagger c_{l \downarrow}^\dagger \ket{0}$ {localized at  single end}{ appearing in $\ket{1_\pm}$,} than for states of type $c_{l \uparrow}^\dagger c_{r \downarrow}^\dagger \ket{0}$ {present} on both ribbon ends. Therefore, the antiferromagnetic state $\ket{0_{S=0}}$, or the state $\ket{0_{S=1}}$ with only one spin species at the {ends}, are the lowest in energy.

{The fundamental gap $E_g$ at half filling is the gap between the {peaks in the} spectral function {due to} electron annihilation and creation [Eq.~(\ref{eq:specfunc})]. {This gap would be seen in STS experiments as the spin-split gap.} {Assuming a frozen core below the {end} orbitals,  $E_g$ at half filling with {$N_\uparrow = N_\downarrow$ particles ($S_z=0$) or $N_\uparrow = N_\downarrow-2$ particles ($S_z=1$),} is}
\begin{align}
E_g& = E^{(N_\uparrow +1,N_\downarrow)}_0 - 2 E^{(N_\uparrow ,N_\downarrow)}_0 + E^{(N_\uparrow,N_\downarrow-1)}_0 \notag\\
&=  U \sum_{\text{sites } i} |\phi_{a}(i)|^2 \left[ |\phi_{a}(i)|^2 + |\phi_{b}(i)|^2 \right] \\
& \overset{\text{large ribbon}}{\longrightarrow} 0.104 \: U.
\end{align}
{The ground states of the half-filled and $(N_\uparrow ,N_\downarrow-1)$-particle systems are characterized by zero {end-end} interaction energies.} In the numerical many-body calculation with $U=2$~eV, the  fundamental gap of the {three-monomer} 7-AGNR is $0.167$~eV, whereas the mean-field model gives $0.256$~eV. In the limit of a long ribbon, the mean-field fundamental gap is $0.254$~eV. The gap {given by} density functional calculations {(0.43~eV) is}, however, much higher, {suggesting} that the Hubbard parameter $U$ should take a larger value in finite-sized ribbons. It should be also noted that the Hubbard model provides only a minimal description of the electron-electron interaction, as the longer-range components of the Coulomb interaction are not explicitly taken into account.}

{The LDOS at the single-particle energies of DFT and the mean-field Hubbard model}, or diagonal elements of the many-body spectral function, are shown in Fig.~\ref{fig:methodcomp} in a simulated $dI/dV$ measurement. The plot on the right shows the energies corresponding to the maps.   In terms of spatial profiles, the three models agree remarkably well, clearly reproducing the same end-localized states close to the Fermi energy. In these states, most of the contribution is from the zigzag end atoms, and the highest amplitude resides on the middle zigzag site. {For the spectral functions of the many-body Hubbard model}, these maps arise directly from creation or annihilation of the end-localized states $\psi_a$ and $\psi_b$. The first bulk states also have similar spatial profiles regardless of the model used. {In contrast to the states close to $E_F$, {their} amplitude {is largest} at the armchair edges, and {they} have little contribution from the zigzag ends}. 

The two lowest exact many-body eigenstates, corresponding to $\ket{0_{S=0}}$ and $\ket{0_{S=1}}$, are close to each other in energy, {and increasing the length of the ribbon decreases the energy gap between them.} Simultaneously, numerical approximations {become less accurate}. Consequently, choosing one {of them} as the initial state for the zero temperature spectral function is questionable. Therefore, the many-body DOS, {i.e.} the trace over the spectral function, is plotted in Fig. \ref{fig:doped_elevels}(a) for the three-monomer ribbon, assuming both states individually as the initial state. At a finite temperature, one would take an average of the two plots, weighted with Boltzmann factors. The spectra have a double-peaked structure at both sides of the Fermi level. This is explained by noting that the lowest eigenstates of the $N\pm1$-particle systems are well-described as single Slater determinants of the tight-binding orbitals. Therefore, for instance the first peaks by annihilation correspond to transitions to states $g^\dagger c^\dagger_{a \uparrow} \ket{0}$ and $g^\dagger c^\dagger_{b \uparrow} \ket{0}$. Since the {end} orbital $\psi_a$ has a slightly lower energy than $\psi_b$, the double peaks are observed. Similar deduction holds on the creation side. Furthermore, the spectra differ in amplitude close to the Fermi level for the case with $\ket{0_{S=0}}$ as {the} initial state, since the weights $\alpha > \beta$ are not equal [see Eq.~\eqref{eq:MB_GS_S0}]. {In longer ribbons, this difference in the height of the peaks cannot be observed, as $\alpha$ and $\beta$ converge to a common value when the ribbon length is increased.}

\begin{figure}
\includegraphics[scale = 0.39]{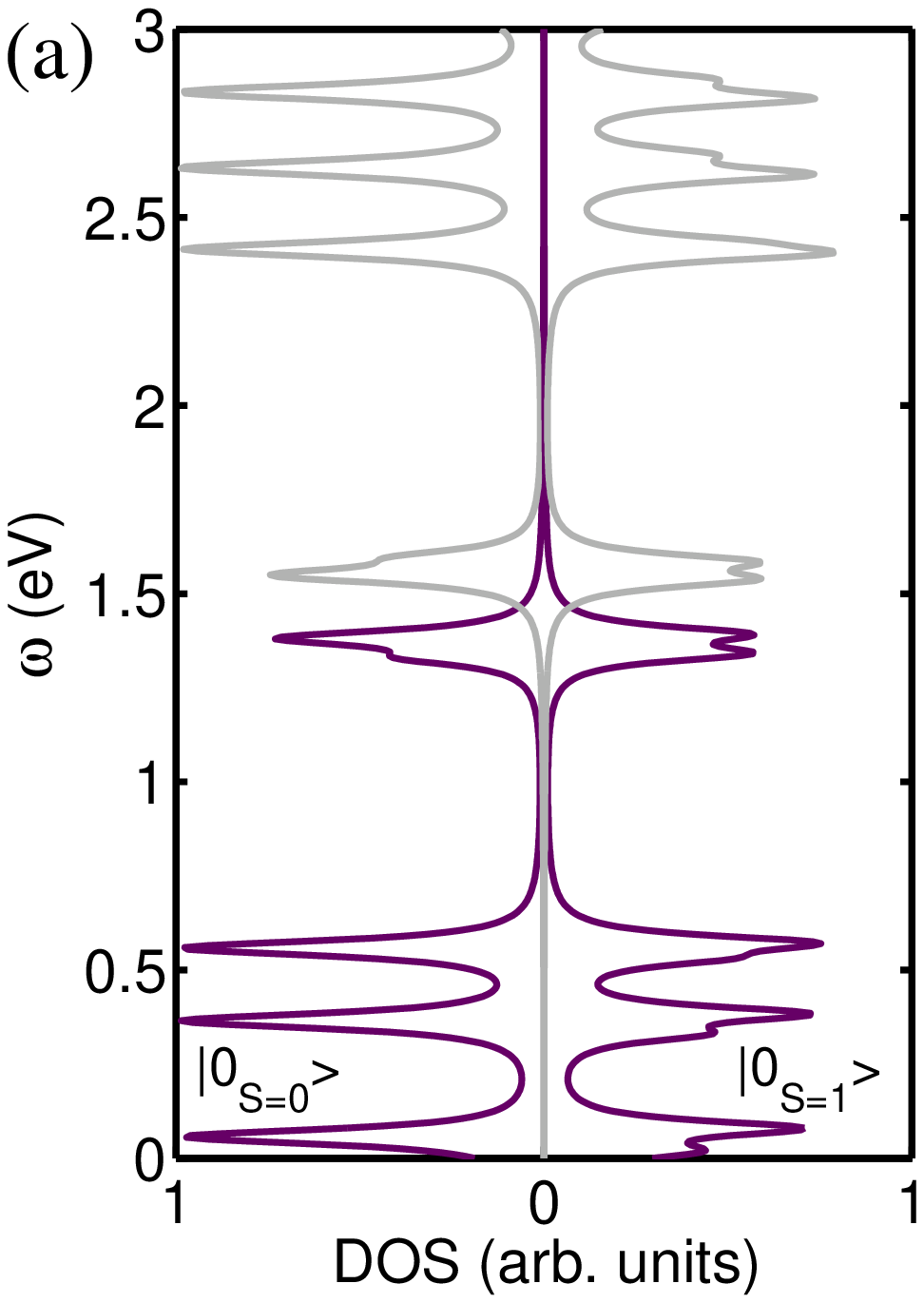}
\includegraphics[scale = 0.39]{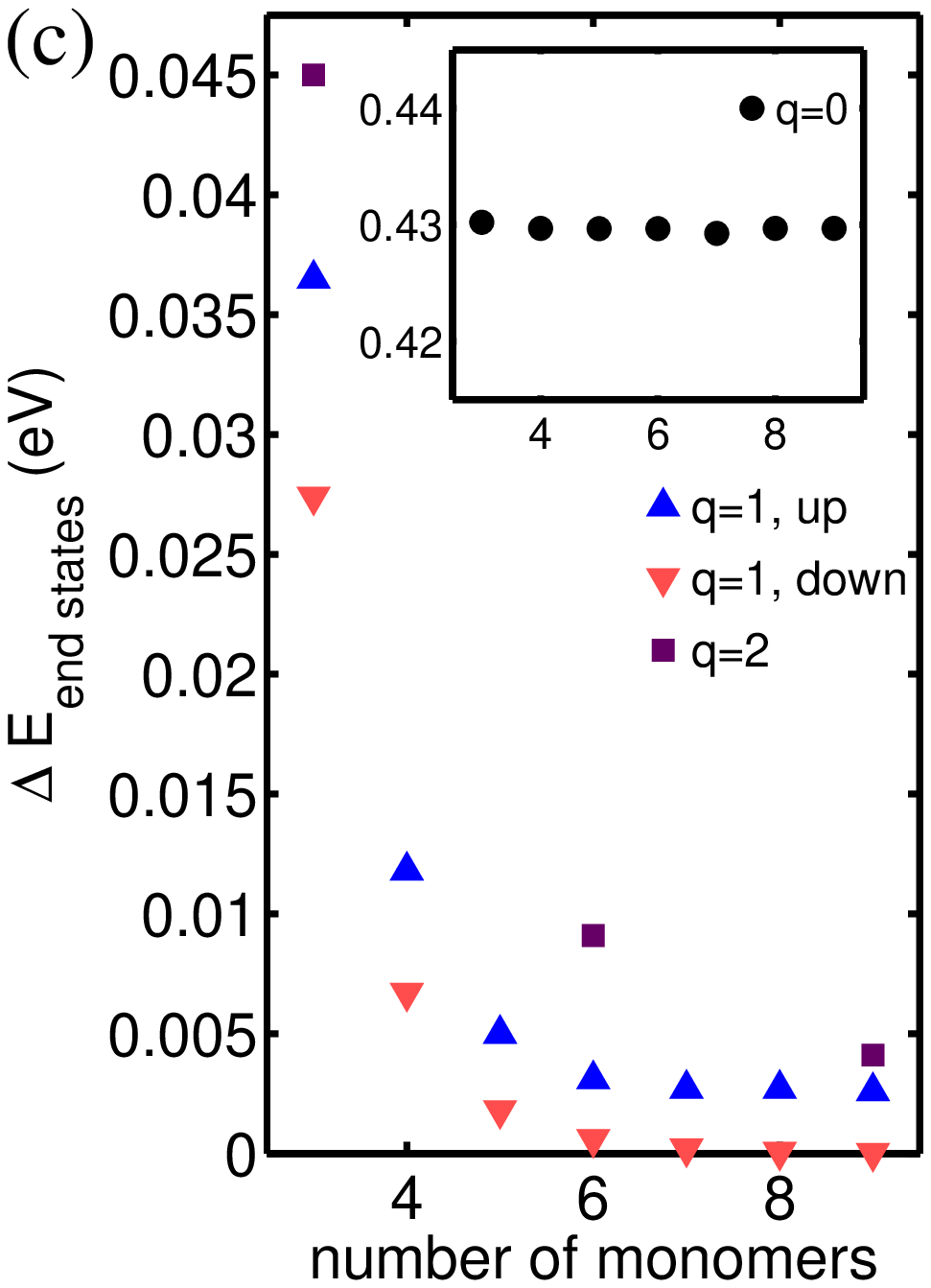}
\includegraphics[scale=0.39]{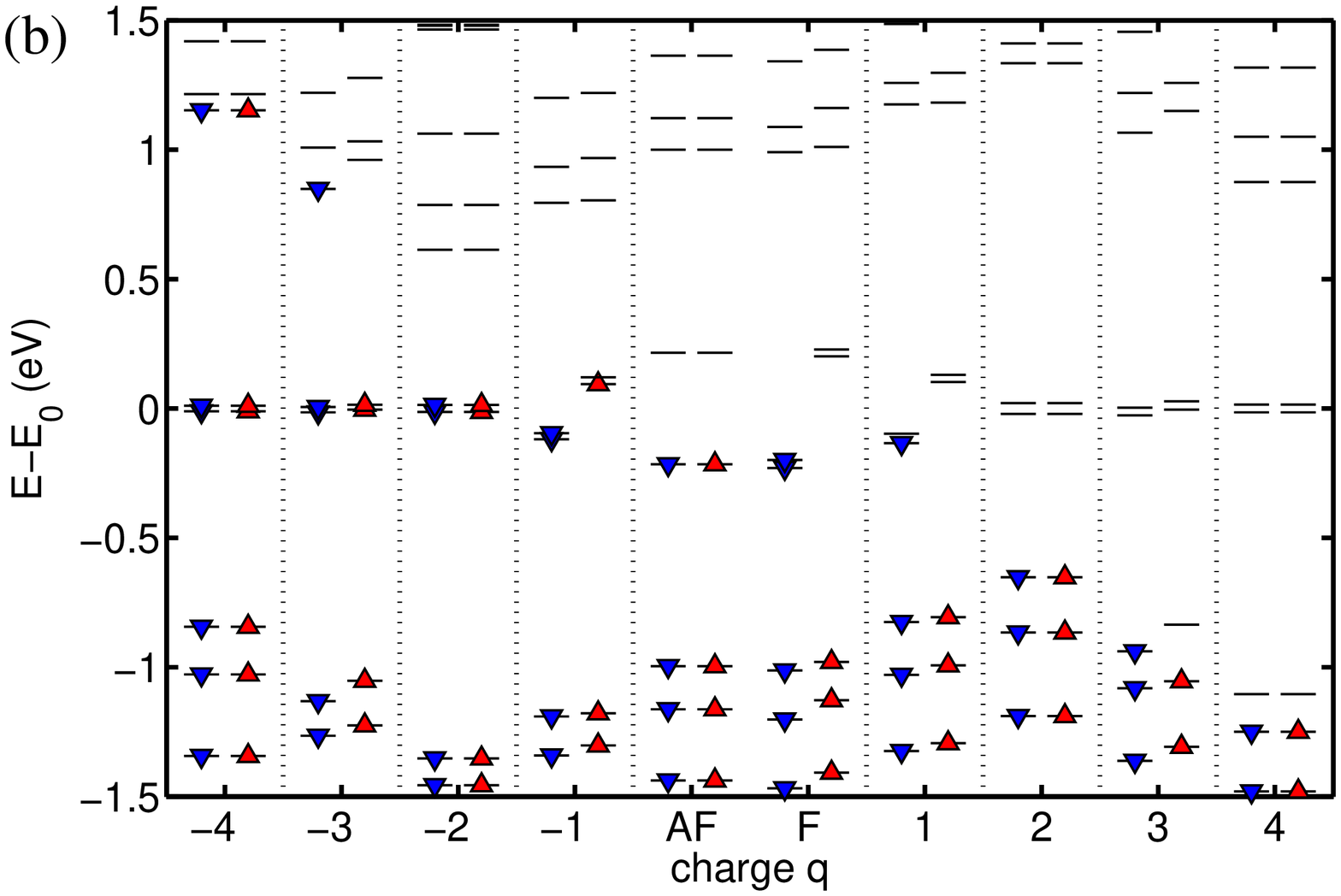}
\includegraphics[scale=0.39]{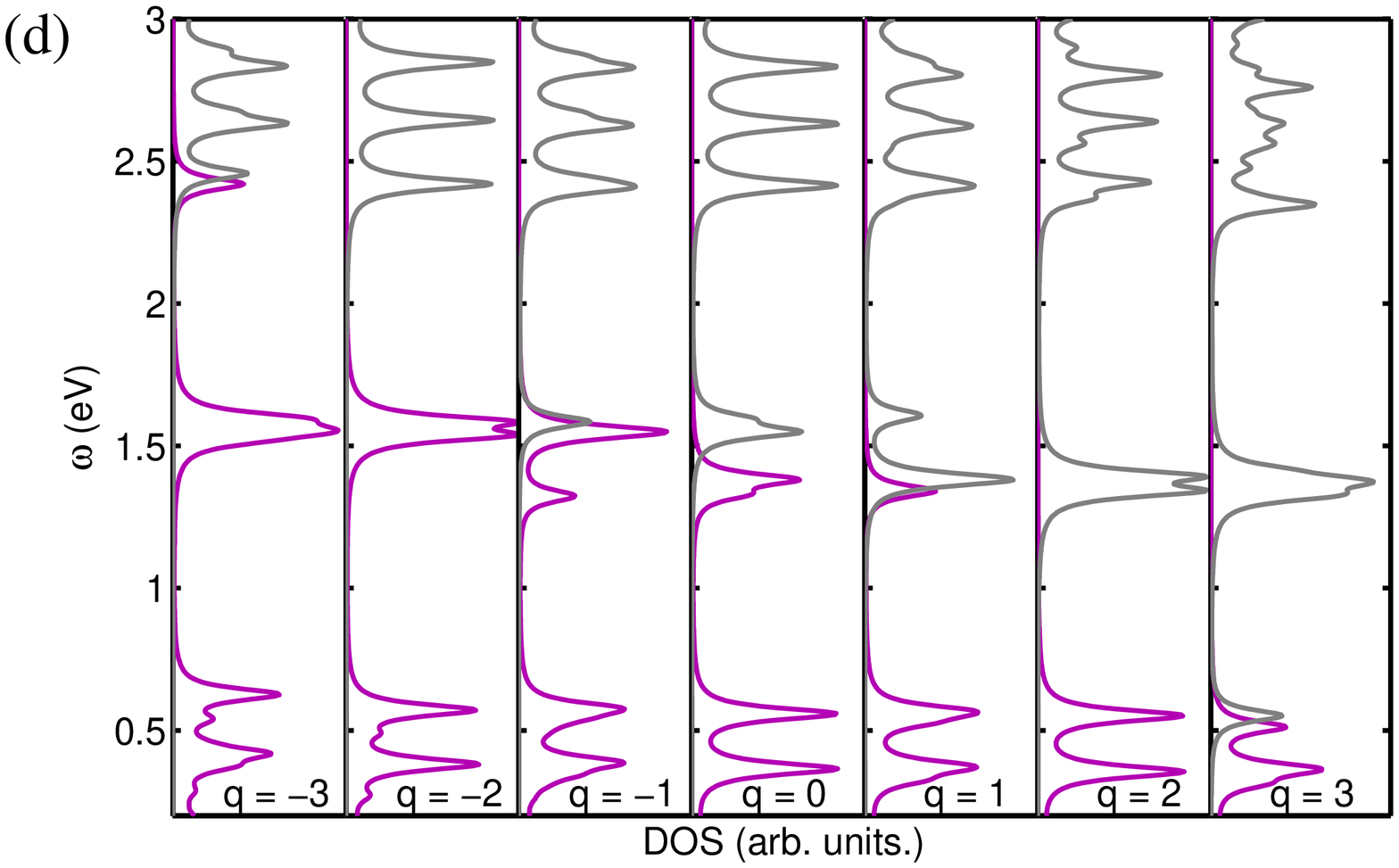}
\caption{\label{fig:doped_elevels} (Color online) (a) Hubbard model many-body density of states at half filling of the three-monomer 7-AGNR. On the left, the initial state is $\ket{0_{S=0}}$, and on the right the initial state is $\ket{0_{S=1}}$ taken as the $S_z=0$ projection. (b) {Kohn-Sham} energy levels of both spin species for the doped three-monomer ribbon. The energy zero ($E_0$) has been normalized to the mid-point of the end-localized states. Lines refer to molecular orbitals, and blue/dark and red/gray triangles refer to the occupied states in the two spin channels. For the uncharged case, the energy levels are shown both for antiferromagnetic (AF) and ferromagnetic (F) coupling between end-localized states. (c) Gap between end-localized states as a function of the number of monomers in the ribbon for the uncharged ($q = 0$, inset) and hole-doped ($q=1,2$) ribbons, {calculated by DFT}. The inset axis labels are the same as in the main figure.  (d) Many-body density of states, now at {various} values of doping with integer excess charge $q$. The purple curves correspond to transitions by annihilation, and the gray curves correspond to transitions by creation of an electron. }
\end{figure}

To summarize our findings comparing the effective one-body models to the full many-body treatment using the Hubbard model, we note that the $dI/dV$ simulations are in {good} agreement. Correlation phenomena are important only in the shell formed from the nearly degenerate end-localized states. DFT, however, predicts a much larger spin-split gap at the zigzag edge {in the uncharged ribbon} than the Hubbard model with $U=2$~eV. 

\subsection{Doped ribbons}

\begin{figure*}
\includegraphics[width = 1.9\columnwidth]{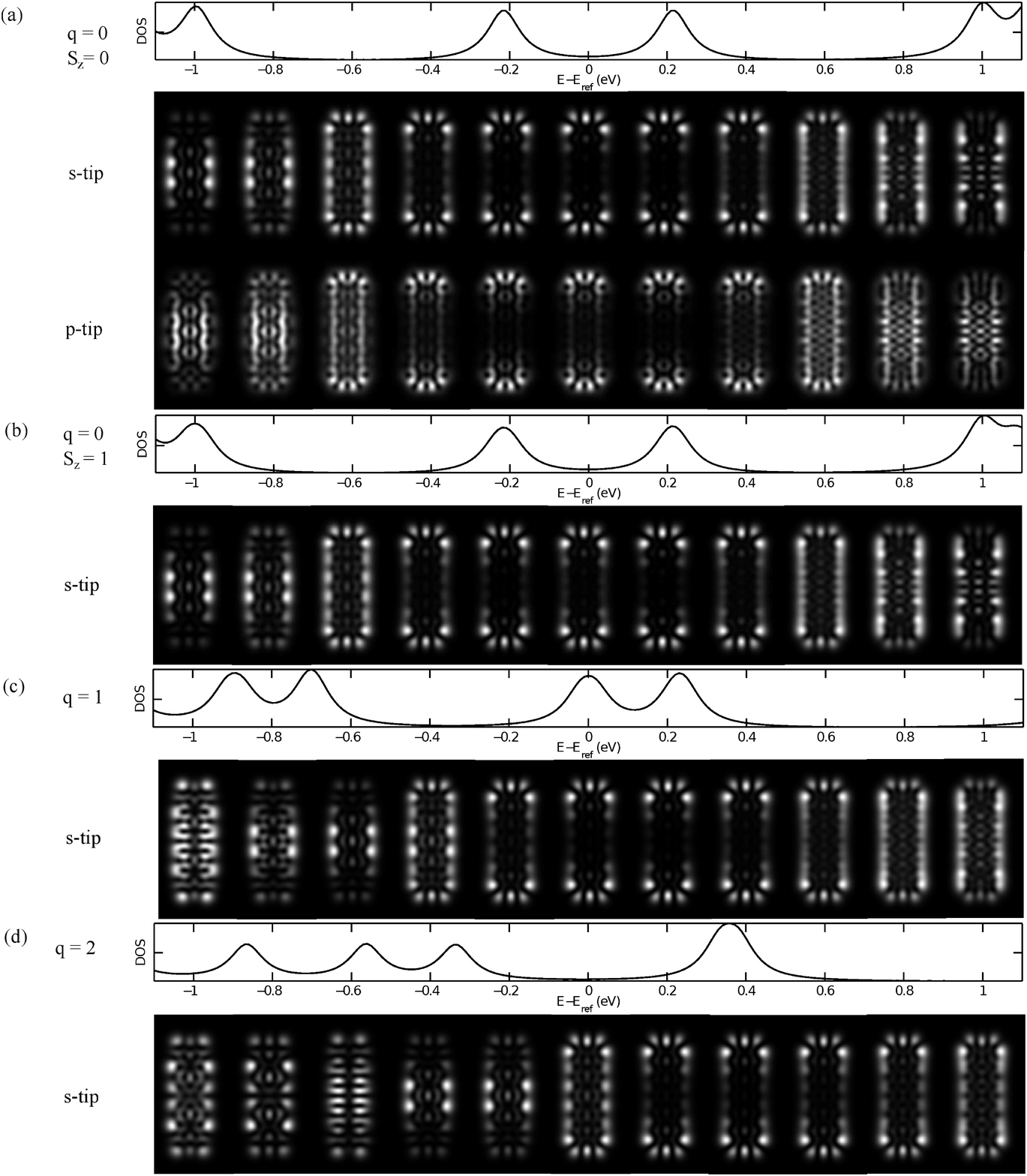}
\caption{\label{fig:fullH} {The DFT-calculated} density of states (DOS) and simulated $dI/dV$ maps for fully hydrogenated ribbons at different values of doping. The energy scale of the DOS figure applies also to the $dI/dV$ maps {in the sense that they} range from -1~eV to 1~eV in steps of 0.2~eV. (a) Antiferromagnetic uncharged system (b) Ferromagnetic uncharged system (c) Hole-doped system with $q$ = 1 (d) Hole-doped system with $q$=2.  The density of states plots show the energies of the molecular orbitals, as well as indicate the overall magnitude of the individual {dI/dV maps}.  }
\end{figure*}

Even in the limit of weak coupling to the substrate, charge transfer between the ribbon and the substrate can occur.\cite{Khomyakov, Repp, Swart} In the experiments,\cite{vanderLit, Koch, Bronner} the zigzag end state peak is observed at positive bias, suggesting that the ribbons are hole-doped. Recently, the interface between Au(111) electrodes and {finite armchair} GNRs {with rounded termini} was studied using DFT, and a charge transfer of {up to} 0.05 electrons per carbon atom {from the ribbon to the electrode} was found.\cite{Archambault} Thus, in finite ribbons, a doping level of few electrons is realistic. The doping of the ribbon might also be non-integer but, for simplicity, we do not consider fractional doping.

{Doping the ribbons has a profound effect on the gap between end-localized states.} The gap {of an} uncharged ribbon {is} constant as a function of ribbon length {both in calculations using DFT and the Hubbard model, see the inset in Fig.~\ref{fig:doped_elevels}(c) for the DFT result}. {This agrees with previous DFT calculations.\cite{Shemella}} In contrast, the gap between the end-localized states {is reduced} by an order of magnitude in hole-doped ribbons ($q=1$) of three monomer units in length, and decreases exponentially with the length of the ribbon.  {The DFT calculation shown in Fig.~\ref{fig:doped_elevels}(c) illustrates this.} The Hubbard model gives the same trend when the ribbon length is increased.  

Due to the presence of additional charge, the HOMO-LUMO gap in doped ribbons does not correspond to the gap between the end states. With an even number of added or removed electrons, the ground state is nonmagnetic with no spin polarization {and all DFT energy levels are spin-degenerate}. Fig.~\ref{fig:doped_elevels}(b) illustrates the {Kohn-Sham} energy level structure of the three-monomer flake as the deviation of the number of electrons from half-filling, {or excess charge} $q$, ranges from four added to four removed electrons. The energy zero has been set to the mid{-point of end-localized state energies}. The gap between the end-localized states in both spin channels is greatly reduced when the ribbon is charged. {With} an odd number of electrons, {for which} $S_z = 1/2$, there is an energy split between the end states belonging to different spin channels.{ It is also worth noting that the position of the end-localized states shift within the bulk gap depending on the {amount of doping}.}

The many-body DOS of the Hubbard model for the three-monomer 7-AGNR is shown in Fig.~\ref{fig:doped_elevels}(d) for integer {values of} excess charge between {$q=-3$ and $q=3$}. The purple lines describe the spectra obtained by annihilation {of electrons from} the initial state, whereas the gray lines describe the spectra obtained by {electron} creation. Th{e DOS} can be directly compared to the corresponding Kohn-Sham energy levels in Fig.~\ref{fig:doped_elevels}(b). Qualitatively, the DFT and many-body DOS plots are surprisingly similar, aside from the half-filled $S=0$ case with the double-peak structure {on} both sides of the Fermi level {in the Hubbard calculation}. {This is no surprise, since the many-body eigenstates for the doped {ribbons}, at least the {few lowest ones}, are well-described in the single particle picture{. F}urthermore, they are close to intuitive excitations from one tight-binding orbital to another.

Fig.~\ref{fig:doped_elevels}(b) shows the molecular orbital energies {at $q=0$} for states with {both} antiferromagnetic (AF) and ferromagnetic (F) coupling between the {ribbon ends}, {i.e.} with $S_z = 0$ and $S_z = 1$, respectively.} In the {ferromagnetic} case, the spin degeneracy of the end-localized states is broken, and a small gap appears between the end-localized states of the same spin, similar to {the case of} general doping. {It is worth noting that these {two DFT} states correspond to {the states} $\ket{0_{S=0}}$ and $\ket{0_{S=1}}$ of the many-body calculation. DFT does not, however, allow one to determine the total spin of a given ground state, only the value of $S_z$. }The energy difference between the AF and F states in the uncharged ribbons decreases when the length of the ribbon is increased, and reduces from the order of~0.001~eV for the three-monomer flake to the limit of numerical accuracy in a nine-monomer flake.  This is in agreement with the Hubbard model results {for $\ket{0_{S=0}}$ and $\ket{0_{S=1}}$ }. 

\begin{figure}
\includegraphics[width=0.95\columnwidth]{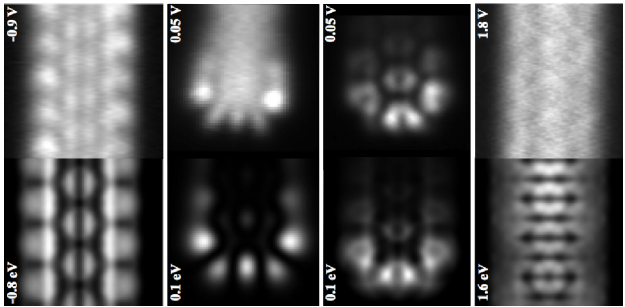}
\caption{\label{fig:exp_theor} {A comparison between experimental $dI/dV$ measurements and DFT simulations. See Ref.~\onlinecite{vanderLit} for experimental details.} Top row: experimental measurements, {measured with a CO-terminated tip, apart from the second column from the left that is measured using a metal-terminated tip.} Bottom row: DFT (PBE functional) simulations, six-monomer fully hydrogenated neutral {ribbon}. A mixed tip with both $s$- and $p$-wave character is used ( $M = 0.15M_s + M_{p_x} + M_{p_y}$), apart from the second column from the left, in which the tip has {pure} $s$-wave character.  The energies corresponding to the maps are indicated in the figure. }
\end{figure}

As the experimental ribbons are rather long, the AF and F states are practically degenerate, and especially at low temperatures, {both} of them {might} be observed. Figs.~\ref{fig:fullH}(a) and (b) show the DFT-simulated DOS and $dI/dV$ maps for the three-monomer ribbon in the AF and F states, respectively, for energies ranging from -1~eV to 1 eV. The maps evaluated between the orbital energies (seen as peaks in the DOS figure) {show a mix of} the {orbitals} due to energy broadening. Fig.~\ref{fig:fullH}(a) also illustrates the difference between $s$- and $p$-wave tips. Areas that are bright in the $s$-wave simulations turn into rings in the $p$-wave simulations. {This effect is demonstrated experimentally in Fig.~\ref{fig:exp_theor}, which compares constant height $dI/dV$ maps of the end-localized state measured with a metal-terminated and carbon monoxide-terminated tips. In addition, $dI/dV$ maps at the onset of the valence and conduction bands measured with a CO terminated tip are shown.}  The simulations correspond well to the experimental measurements, if a tip with mixed $s$- and $p$-wave character is assumed, with the tunneling matrix element given by $M = 0.15M_s + M_{p_x} + M_{p_y}$. \footnote{The magnitude of the $s$- and $p$-wave contributions vary spatially.\cite{Gross, Pavlicek} The relative tip contributions in the simulations should thus not be used directly to characterize the experimental tip. }

As seen in Figs.~\ref{fig:fullH}(a) and (b), the $dI/dV$-maps for the F and AF states as well as the DOS plots are practically indistinguishable. In addition, the energies of the occupied and unoccupied states are similar.  Consequently, the AF and F states cannot be distinguished with STM using non-magnetic tips. {The only noticeable difference is in the peak amplitudes due to the many-body nature of the states, but this could be hard to pinpoint clearly in an experiment.} {{STM experiments using a spin-polarized tip or measurements performed in a magnetic field} could possibly be used to distinguish between the magnetic states.}

The effect of doping on the DFT-calculated $dI/dV$ maps is illustrated in Fig.~\ref{fig:fullH}(c) and \ref{fig:fullH}(d) for $q=1$ and $q=2$, respectively.  Even though the energy level spacings are clearly changed upon doping,  the orbitals remain spatially similar. The main effect of doping on the $dI/dV$ maps is a shift of the energy axis{. T}he end state resonance mov{es} to the occupied side of the spectrum {as} the amount of hole-doping increases. 

The amount of doping required to quench the {antiferromagnetic coupling between} the ends of a finite AGNR{, and the associated spin-split gap,} has not been previously addressed. Kunstmann~\emph{et al.}\cite{Kunstmann} found that for an infinite 12-ZGNR, whose width corresponds to the length of the three-monomer ribbon, a doping level of 0.5 electrons per edge atom was enough to destroy the spin moments. This corresponds to {a doping of} three electrons in the finite ribbons.  {Our calculations indicate that doping by even a single charge carrier is enough to quench the antiferromagnetic order.}

\subsection{Effect of defects}

{Despite {describing} only the graphene $\pi$-electrons, the Hubbard model seems to at least qualitatively agree with the DFT results. Being computationally inexpensive, the mean-field approximation allows one to reach large system sizes. {The neglect of the $s$- and $\sigma$-electrons in the lattice approach makes this method in principle less well suited to study defects affecting also the carbon atom hybridization, or coordination. Simple defects, however, can be modeled in a crude way, and below, we will compare mean-field Hubbard calculations to DFT results to study the usefulness of this approach. }

Fig.~\ref{fig:def_dft_mf} compares the simulated $dI/dV$ spectra from DFT and {the} mean-field Hubbard {model} for a three-monomer ribbon with a {single} CH$_2$ group at the armchair edge, modeled as an edge vacancy in the Hubbard model.  The structure of the defected ribbon is shown in Fig.~\ref{fig:defect_structures}(a). Even though  the vacancy description qualitatively reproduces the defect-localized state of the DFT calculation{, the nodal plane structure in the vicinity of the defect is missing from the map calculated using the mean-field Hubbard model}. {In the DFT-calculated map, for instance, there is some amplitude on the added hydrogen atom,} whereas the lattice model gives most weight to the dangling carbon site at the armchair edge next to the defect {and naturally none to the defect site}.  It is thus clear that such a simplified description is unable to capture all characteristics and, consequently, in further studies on {defects and edge hydrogenation}, we only use DFT. }

\begin{figure}
\includegraphics[width=0.95\columnwidth]{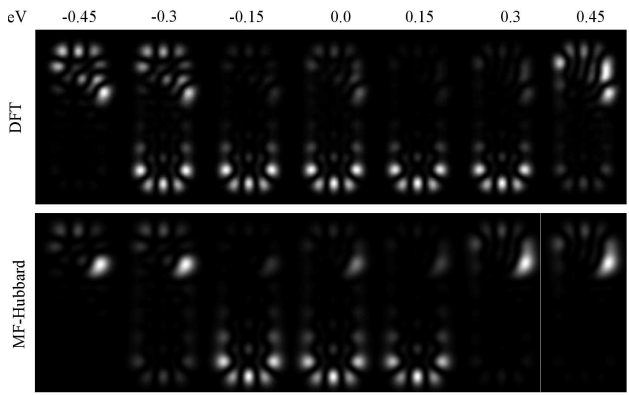}
\caption{\label{fig:def_dft_mf} A comparison between DFT (top row) and the mean-field Hubbard model (bottom row) in modeling the $dI/dV$ maps of a three-monomer ribbon with a CH$_2$ group at the armchair edge on the fourth armchair carbon atom [see Fig.~\ref{fig:defect_structures}(a) {for the structure}]. }
\end{figure}

\begin{figure}
\includegraphics[width = 0.85\columnwidth]{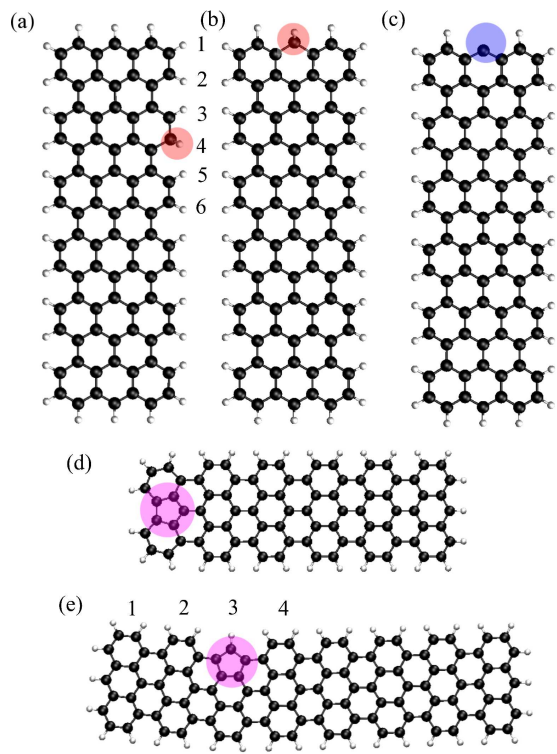}
\caption{\label{fig:defect_structures} (Color online) Considered structures for defected finite ribbons.  Different possible positions for a defect type have been marked using numbers.  (a) CH$_2$ group at the fourth armchair edge atom [CH2(4)] (b) CH$_2$ group at ribbon end [CH2] (c) Asymmetrically missing hydrogen atom at the middle zigzag carbon. [nmH/H]  (d)  Missing CH group at the end [penta] (e) Missing CH at the armchair edge, at the third armchair dent [penta(3)]. {The colored circles highlight the positions of the defects (missing hydrogen atom: blue, CH$_2$ group: red, pentagon: purple).}}
\end{figure}

\begin{figure}
\includegraphics[width = 0.9\columnwidth]{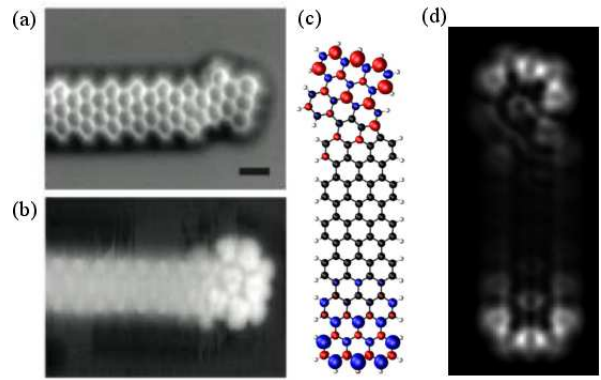}
\caption{\label{fig:expdefect} (Color online) Ribbon with a misplaced monomer at one terminus. (a) Experimental AFM scan showing the single pentagon in the carbon backbone. (b) Experimental {STM topographic map (10mV, 100pA)} {Both experimental images have been obtained using a carbon monoxide terminated tip. For experimental details, see Ref.~\onlinecite{vanderLit}.} (c) Structure and spin density {given by} DFT. Maximal spin polarizations: intact end $\mu$ = 0.198~$\mu_B$, defected end: $\mu$ = 0.182~$\mu_B$.  (d) DFT simulation of the low bias $dI/dV$ map at 0~eV, using a mixed $s$- and $p$-wave tip ( $M = 0.15M_s + M_{p_x} + M_{p_y}$). }
\end{figure}

\begin{figure*}
\includegraphics[width = 1.95\columnwidth]{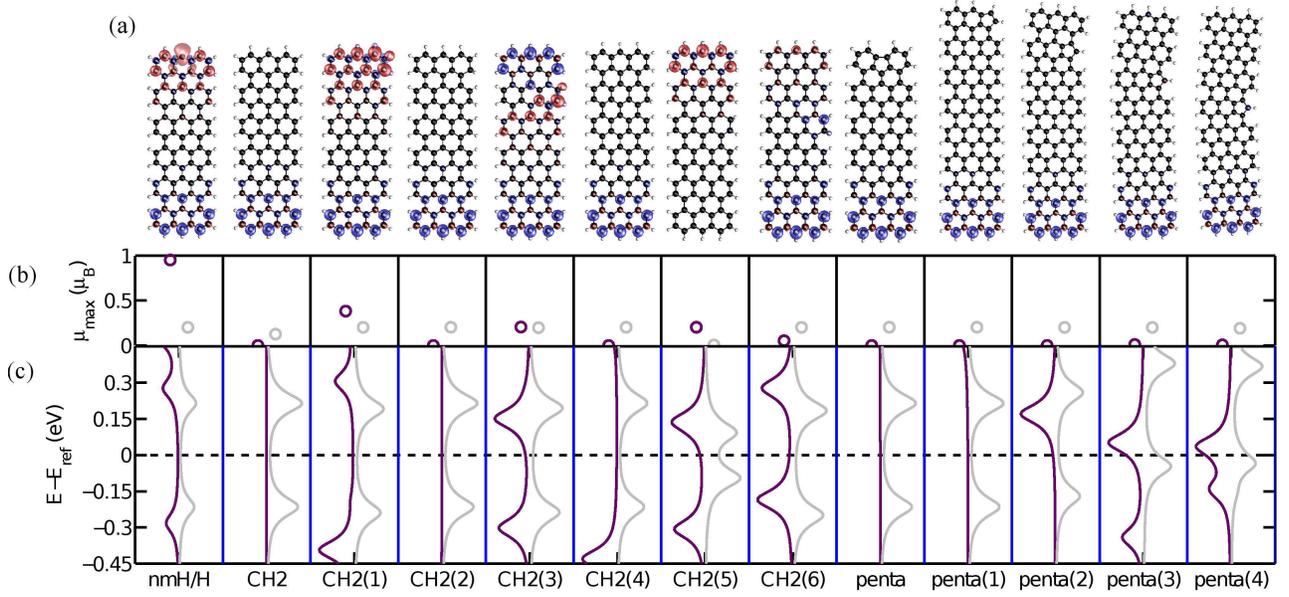}
\caption{\label{fig:defect_mu_splits} (Color online) {Defected ribbons} {(a)} The structure and the spin density  for both spin species (red/gray and blue/dark, isosurface value 0.01~e/\AA{}$^3$). {(b)} The maximal spin moment $\mu_{\mathrm{max}}$  {(c)} The local density of states in the middle of the zigzag termini. {In (b) and (c), the colors refer to the ribbon ends.} Purple/dark-- the end closer to the defect,  gray -- the intact end. } 
\end{figure*}

\begin{figure}
\includegraphics[width = 0.98\columnwidth]{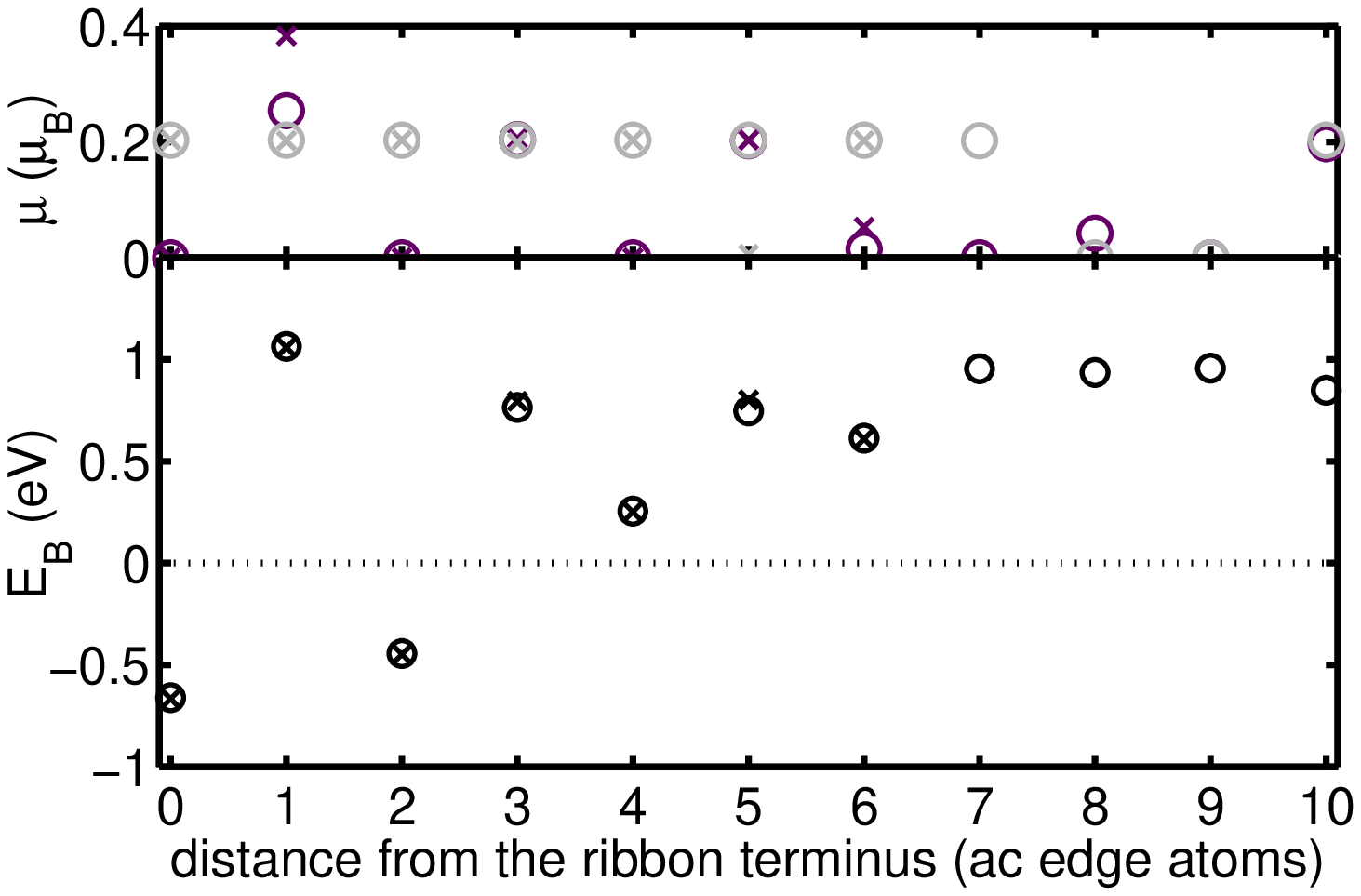}
\caption{\label{fig:EB_CH2} (Color online) The {maximal spin} moment $|\mu|$ and the binding energy at the defected end (purple/dark) and the intact end (gray) as a function of the distance of the {CH$_2$} defect from the ribbon termini. Crosses -- three-monomer ribbon, circles -- five-monomer ribbon }
\end{figure}

We will consider two main classes of defects. The first kind,  additional hydrogen atoms at the edge forming CH$_2$ groups {[Fig.~\ref{fig:defect_structures}(a) and \ref{fig:defect_structures}(b)]}, has been observed experimentally.\cite{vanderLit, Talirz}  In Refs.~\onlinecite{vanderLit, Talirz}, the appearance of the end-localized states could be modified by dehydrogenating the flake ends. Similarly, defects can be introduced at the armchair edge. {Thus}, we include a ribbon with a removed middle zigzag hydrogen at one terminus  {[Fig.~\ref{fig:defect_structures}(c)]}. The {last defect type} mixes the sublattices by removing a CH group, forming thus a pentagon at the ribbon {end or} edge {[Fig.~\ref{fig:defect_structures}(d) and \ref{fig:defect_structures}(d)]}.  Little is known about how such structural defects affect the {electronic structure of finite ribbons}. The effect of the defect position is studied both for a CH$_2$ group and a pentagon at the ribbon edge. The distance of the defect from the ribbon end is measured in units of armchair edge carbon {atoms}. 

{Some structural defects can be unambiguously identified using AFM measurements. Fig.~\ref{fig:expdefect} shows experimental AFM and $dI/dV$ measurements of a ribbon with a tilted terminating monomer unit, as well as the corresponding theoretical calculations of the spin density, {and} a $dI/dV$ map of the end states. For details on the nanoribbon synthesis as well as the STM and AFM measurements, please refer to Ref.~\onlinecite{vanderLit}. } {$dI/dV$ fingerprints may aid in identifying various defect types, especially if AFM maps are not available.}  

\begin{figure}
\includegraphics[width =\columnwidth]{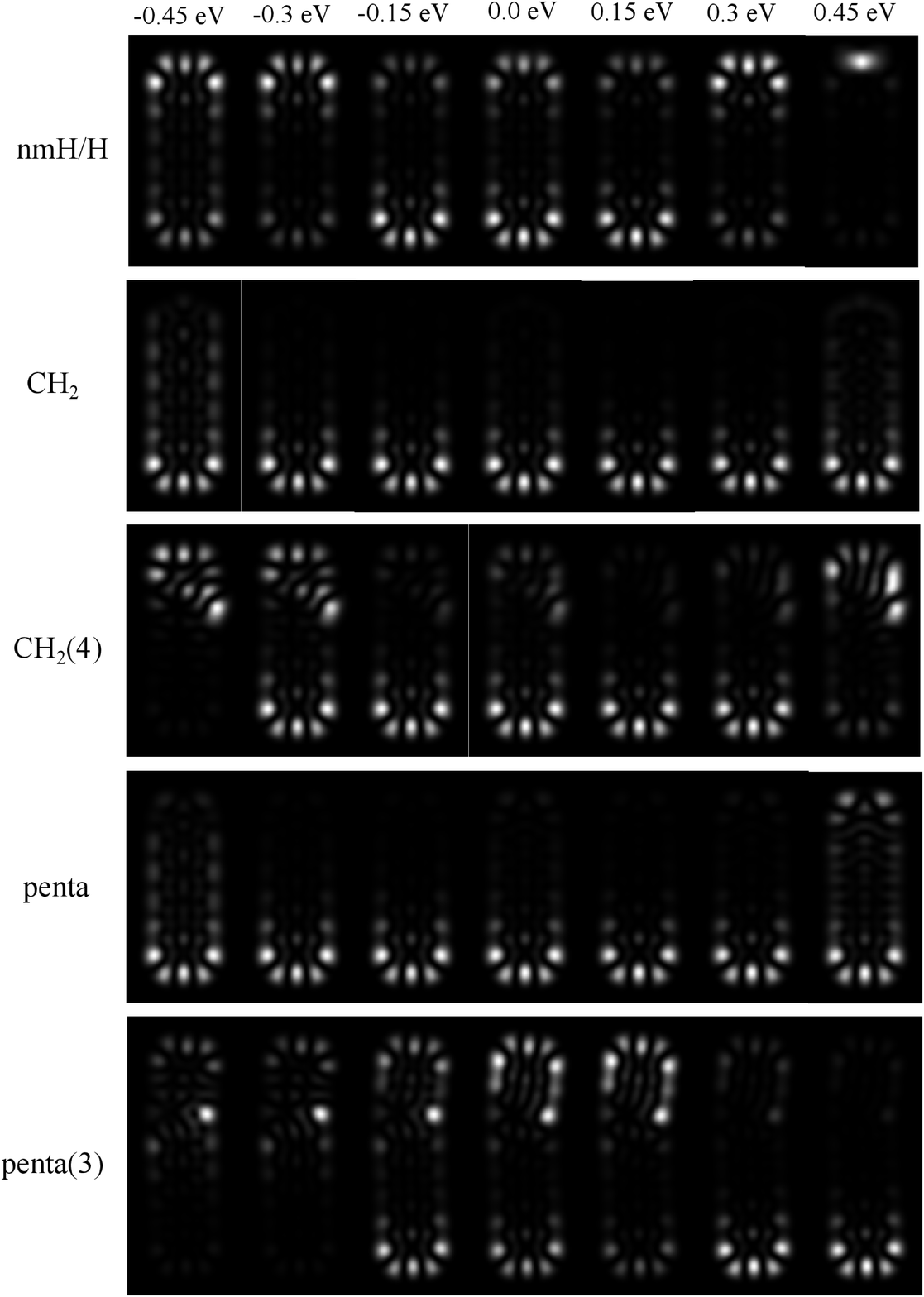}
\caption{\label{fig:defect_stm} $s$-wave $dI/dV$ map simulations for some of the defected ribbons. The top end is closer to the defect. For an illustration on the structures and naming conventions, see Fig.~\ref{fig:defect_structures}. }
\end{figure}

Fig.~\ref{fig:defect_mu_splits}(a) shows for different defected structures the spin densities {in red/gray and blue/dark for the two spin species}, and Fig.~\ref{fig:defect_mu_splits}(b) the maximal spin moments $\mu_{\mathrm{max}}$ at both the defected end (purple/dark) and at the intact end (gray) for the different defect structures.  Apart from few exceptions, the maximal moment at the intact end is little affected, and $\mu_{\mathrm{max}} \approx 0.2 \mu_B$ is very close to the value of a pristine, fully hydrogenated ribbon {($\mu_{\mathrm{max}} = 0.202 \mu_B$)}. At the defected end, however, the behavior depends both on the defect type and its distance from the ribbon end. Removing the middle zigzag hydrogen atom only from one end (nmH/H) leads to an electron almost fully localized at the defect site, with a spin moment of 0.95~$\mu_B${, assuming no coupling with the substrate}. On the other hand, a CH$_2$ (named CH2 in Fig.~\ref{fig:defect_mu_splits}) group or missing CH group (named penta) at the zigzag end completely quenches the spin polarization at the affected end. As the corresponding $\pi$ electron is removed from the $\pi$ system, the sublattice-alternating pattern of the end state is no longer possible.  When the CH$_2$ defect is moved along the armchair edge farther away from the ribbon end, an oscillating behavior is observed as a function of the distance from the terminus {[Fig.~\ref{fig:defect_mu_splits}(b)]} The spin moment is quenched if the CH$_2$ group is attached to a carbon atom belonging to the same sublattice as the zigzag {end} atoms in the closer ribbon termini. {Again, the quenching of the end state is related to the elimination of a site with a large contribution to the zigzag end state.} 

In order to better understand the oscillating behavior caused by the CH$_2$ defect, Fig.~\ref{fig:EB_CH2} compares the maximum spin moments at the defected (purple/dark) and intact end (gray), as well as the binding energy of the second added hydrogen atom to the different edge sites, defined as 
 \begin{equation} E_B = E_{\mathrm{ribbon+H}}-E_{\mathrm{ribbon}}-\frac{1}{2}E_{\mathrm{H}_2}.
\end{equation}
{A positive $E_B$ thus implies that the binding of a second hydrogen atom to a site is energetically unfavorable. } Results for both three- and five-monomer ribbons are shown in Fig.~\ref{fig:EB_CH2}, marked with crosses and circles, respectively. The H binding energy $E_B$ is found to increase with the distance from the terminus, {superimposed with the sublattice-dependent oscillation}, {and} the attachment of a hydrogen atom {becomes more unfavorable}. $E_B$ is smaller for sites belonging to the same sublattice as the zigzag end state, and thus quenching the spin moment at the defected end is actually energetically favored. In the case of quenching, the local density of states at the defected end [Fig.~\ref{fig:defect_mu_splits}(c)] does not show end-localized molecular orbitals {close to $E_F$}. For defects attached to a carbon atom belonging to the other sublattice, the magnetic moment at the terminus closer to the defect decreases {as a function of distance} toward the value in pristine ribbons.

Defects located far away from the terminus in a five-monomer ribbon, {more precisely} at locations 8 and 9, {form} a localized state at the armchair edge that quenches the zigzag end state at both termini. {For some defect positions, multiple magnetic states are possible. For instance in the case of the CH$_2$(3) defect [Fig.~\ref{fig:defect_mu_splits}], the ribbon termini may show either AF or F order {with respect to each other}. With AF order, the state  localized on the defect may correspond to either of the spin species. The energy differences between the different magnetic states are of the order of tens of meV, and {the magnetic state has {only a minor} effect on} the LDOS curves.} 

A pentagon defect at the armchair edge always quenches the spin moments at the end closer to the defect, as the end-localized states are pushed up in energy {to the unoccupied side of the spectrum.} The electronic structure at the intact end is almost unchanged [Fig.~\ref{fig:defect_mu_splits}], and the energy split between the end-localized states remains roughly constant at 0.4~eV. The states may, however, shift with respect to the HOMO and LUMO of the defected system. 

Fig.~\ref{fig:defect_stm} shows simulated $s$-wave tip maps for  defected ribbons with different defect types. Depending on the defect type, the zigzag end state is observed only at one end, or at both ends at different bias {voltages}.  {A CH$_2$ group at the ribbon end completely quenches the zigzag end state, as observed experimentally.\cite{Talirz, vanderLit} Both CH$_2$ and pentagon defects at the edge show a state localized at the defect but with different nodal patterns. }{This suggests that, in addition to AFM, defects can be identified using $dI/dV$ imaging. }

\subsection{Different edge hydrogenation patterns}

\begin{figure*}
\includegraphics[width = 1.9\columnwidth]{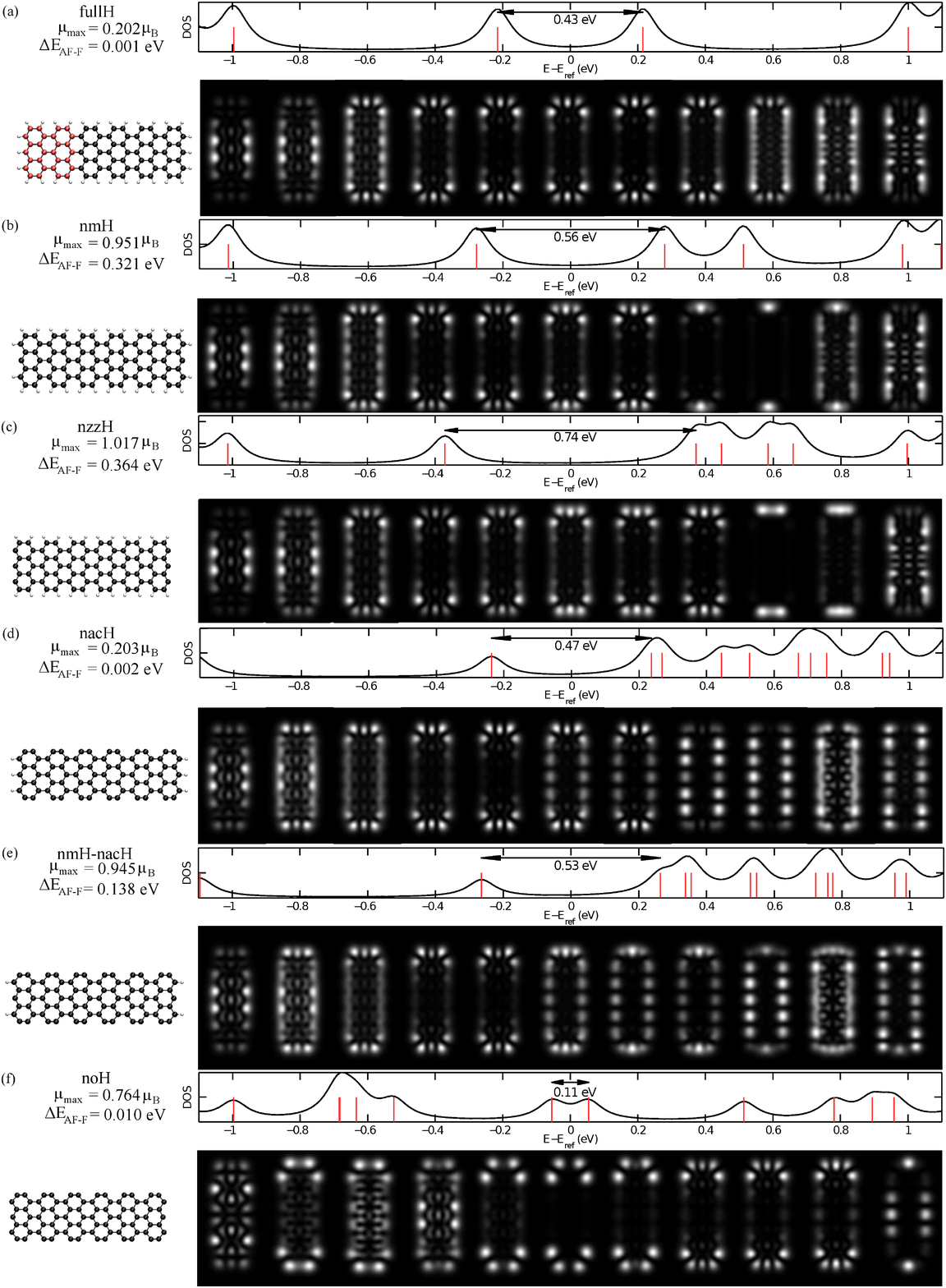}
\caption{\label{fig:edgeterm} \footnotesize (Color online) {DFT-}calculated density of states (DOS) showing the energies of the molecular orbitals, marked using red/gray lines, as well as simulated $s$-wave $dI/dV$ maps for three-monomer undoped 7-AGNRs with different edge terminations. {{The energy scale of the DOS figure applies also to the $dI/dV$ maps {in the sense that they} range from -1~eV to 1~eV in steps of 0.2~eV.} Negative and positive energies correspond to occupied and unoccupied states, respectively. } Structural images are shown on the left [red atoms in (a) mark a structural monomer], and the magnitude of the HOMO-LUMO gap is indicated with a double arrow in the DOS panel. The maximal spin polarization $\mu_{\mathrm{max}}$ and the energy difference between antiferromagnetically and ferromagnetically ordered end states, $\Delta E_{\mathrm{AF-F}}$, is shown for each structure. From top to bottom: full hydrogenation (fullH),  missing H in the middle zigzag site (nmH), missing H at zigzag edge (nzzH), missing armchair H (nacH), missing middle zigzag H and armchair H (nmH-nacH),  and completely dehydrogenated structure (noH).   }
\end{figure*}

We will {now} address the effect of different edge terminations on the electronic structure {and end-localized states by considering} a {three-monomer} ribbon. The structures are shown on the left in Fig.~\ref{fig:edgeterm}. In addition to the fully hydrogenated ribbon (fullH) [Fig.~\ref{fig:edgeterm}(a)], we compare $dI/dV$ simulations for a structure having no middle zigzag hydrogen {atoms} at {either} end, thus resembling the monomer structure after the detachment of a bromine atom (nmH) [Fig.~\ref{fig:edgeterm}(b)], no hydrogen atoms at the zigzag edge (nzzH) [Fig.~\ref{fig:edgeterm}(c)], no hydrogen atoms at the armchair-terminated edge (nacH) [Fig.~\ref{fig:edgeterm}(d)], no hydrogen atom at the middle zigzag carbon or at the armchair edges (nmH-nacH) [Fig.~\ref{fig:edgeterm}(e)],  and a structure completely stripped of terminating hydrogen atoms (noH) [Fig.~\ref{fig:edgeterm}(f)]. One could also consider a structure in which the precursor bromine atoms have not been detached from the ribbon termini. We have excluded this structure from our analysis as it has been considered in Ref.~\onlinecite{Koch}. We first discuss the bulk states, {which have} the largest contribution {to the LDOS at} the armchair edges, before moving to the zigzag end states. 

On the occupied side of the spectrum (at negative energies), the LDOS as well as the simulated $dI/dV$ maps are rather similar regardless of the edge termination for the bulk states, as seen in Fig.~\ref{fig:edgeterm}. On the unoccupied side (at positive energies), on the other hand, the simulated $dI/dV$ maps show marked differences between the edge terminations. Firstly, a hydrogen atom missing at the middle zigzag carbon manifests itself as a strongly localized state {around 0.5~eV as seen in Figs.~\ref{fig:edgeterm}(b) and \ref{fig:edgeterm}(e)}. Secondly, the absence of armchair hydrogen atoms leads to states localized at the armchair edge, unambiguously identifiable as a row of strong resonances in the $s$-wave image [nacH, nmH-nacH, and noH, Fig.~\ref{fig:edgeterm}(d), \ref{fig:edgeterm}(e) and \ref{fig:edgeterm}(f), respectively]. A similar localized edge resonance is seen at the dehydrogenated zigzag edge (nzzH) but{, in this case,} extending to all three zigzag carbon atoms. Dehydrogenated carbon atoms at the zigzag and armchair edges can be expected to be reactive. For ribbons on surfaces, this will lead to bond formation with the substrate\cite{vanderLit} and thus possibly to the suppression of the armchair edge states. In the case of infinite zigzag ribbons, binding between unhydrogenated edge carbon atoms and Au(111) has been predicted.\cite{Li-Zhang} This {causes curving of the ribbon and} quenches the edge magnetization.

Comparing the simulated $dI/dV$ maps to the experimental maps shown in Refs.~\onlinecite{vanderLit, Koch}, we conclude that it is unlikely that the armchair edge hydrogen atoms are missing, as the typical row of strong resonances for structures without armchair-H has not been observed in experiments. {{Moreover}, the dehydrogenated edge {would probably} bind to the substrate.} Also a missing zigzag hydrogen atom should be observable as a bright, localized resonance. When comparing the simulations with experiment, one needs to remember that for molecular systems, the experimental peak positions also contain contribution from the charging energy that is missing from the picture based on the Kohn-Sham eigenvalues. Moreover, the unoccupied DFT can at most be treated as a rough approximation to the measurements at positive voltage. 

{We now turn to the states localized at the zigzag termini of the ribbons.} Fig.~\ref{fig:edgeterm} shows that the spatial profile of the zigzag end state in the $dI/dV$-maps around zero energy is similar regardless of the end-terminating hydrogen atoms. This state is characterized by five bright lobes at the carbon atoms belonging to the same sublattice as the terminating zigzag carbons.  The noH structure is an exception as around zero energy, only four lobes are seen, and the five-lobe edge state occurs at  around 0.6~eV.  

The gap between the HOMO and the LUMO, corresponding to the end-localized zigzag edge states in all structures but noH, is indicated with double arrows in the DOS plot in Fig.~\ref{fig:edgeterm}. The removal of the middle zigzag hydrogen atom or the armchair hydrogens atoms does not strongly modify th{is gap} (the gap between end-localized states), which is around 0.5~eV in the fullH, nmH, nmH-nacH and nacH structures. In contrast, the removal of all zigzag hydrogen atoms increases the gap to 0.74~eV when the armchair hydrogenation is intact, and decreases the HOMO-LUMO gap {of the four-lobe states} to 0.11~eV for the completely dehydrogenated structure. In the noH structure, the HOMO and the LUMO form a second pair of end-localized states, in addition to the five-lobed end state{, with an energy split of 0.27~eV around 0.6~eV}.  It is worth noting that the gap between end-localized states is independent of the number of monomers in the ribbon, see the {inset} in Fig.~\ref{fig:doped_elevels}(a). This is dissimilar to periodic zigzag nanoribbons, in which the split is roughly inversely proportional to the ribbon width.\cite{Son}

{It is well-known that {the PBE-functional} underestimates band gaps, and thus {the calculated} bulk states are at lower energies on the unoccupied side than expected based on the experimental{ly determined} band gap (2.3-5.1~eV).\cite{Bronner,Koch,Linden,Ruffieux} Using the PBE functional, the onset of the occupied and unoccupied bulk states occurs at -0.8 to -1.1~eV and at 0.3 to 1.0~eV, respectively{, for the different edge terminations}. The use of the B3LYP hybrid functional increases the energy level spacing. In the fullH structure,  the end-state gap and the bulk gap increase from 0.43~eV to 1.46~eV, and from 2.00~eV to 2.94~eV, respectively, {when the B3LYP functional is used}. Thus, the spacing between localized states increases more than that of extended states but the simulated $dI/dV$ maps remain, however, qualitatively similar. The {larger shift} for more localized states is also seen in the nmH structure. The localized resonance at the site with a missing hydrogen atom is pushed to higher energy with respect to the first bulk state, and the gap between the localized state and the onset of bulk states decreases from 0.47~eV to 0.20~eV for PBE and B3LYP, respectively.}  Our results{, acquired} using a hybrid functional, are in line with previous investigations. Hod~\emph{et al.}\cite{Hod-Peralta-Scuseria} found a spin split of 1.38~eV between the end-localized zigzag states in {finite} a 9-AGNR. 

\section{Conclusions}

We have studied the electronic states localized at the zigzag termini of finite 7-AGNRs using the Hubbard model and density-functional theory. {We have shown that the $dI/dV$ maps calculated with DFT and the Hubbard model match qualitatively. Furthermore, correlation phenomena are only relevant in the {nearly degenerate shell formed by the} end-localized states of uncharged ribbons, as the states of the charged ribbons can be well described in the single particle picture. }

By studying both hole- and electron-doped ribbons, we have shown that the energy gap between the end-localized states is greatly reduced as compared to neutral ribbons. Furthermore, the spacing between the experimentally observable $dI/dV$ peaks decreases to the order of a few meV, and {it also} decreases as a function of ribbon length. In $dI/dV$ maps, a {charged} state of a ribbon manifests itself mostly as a shift of the energy axis. {Our results indicate that the experimentally observed double peak\cite{vanderLit} at low positive bias in the STS spectrum measured at the ribbon ends cannot arise from two distinct electronic states. }By considering {ribbons with defects at one end}, we provide further evidence for this interpretation. 

In the experiment\cite{vanderLit}, modifying one ribbon end leads to suppression of the double peak located at higher energy at the intact end. We found, however, little changes in the electronic structure of the intact end{, when the other end was modified with a CH$_2$ group, a pentagon defect or a missing hydrogen atom}. Finally, we have calculated $dI/dV$ maps for ribbons with different edge hydrogen terminations, which may aid in the interpretation of STS experiments. 

\acknowledgements
M.I. acknowledges financial support from the V{\"{a}}is{\"{a}}l{\"{a}} foundation and from Finnish Doctoral Programme in Computational Sciences FICS. This research has also been supported by the Academy of Finland through its Centres of Excellence Program (project{s} No. 251748 {and No. 250280}){, NWO (Chemical Sciences, Veni-grant 722.011.007), as well as by European Research Council (ERC-2011-StG 278698-PRECISE-NANO)}. We acknowledge the computational resources provided by Aalto Science-IT project and Finlands IT Center for Science (CSC).

\bibliography{spin_split_with_Hubbard_v4}

\end{document}